\documentclass[apj]{emulateapj}
\usepackage{tabularx}
\usepackage{multirow}
\usepackage{color}
\usepackage{upgreek}
\definecolor{naviblue}{RGB}{0,0,102}

\shorttitle{X-ray/radio observation of MWC~656}
\shortauthors{Rib\'o et al.}

\begin{document}

\title{The First Simultaneous X-Ray/Radio Detection of the First B\lowercase{e}/BH System MWC~656}

\author{M. Rib\'o$^{1}$\footnote{Serra H\'unter Fellow.}, P. Munar-Adrover$^{2}$, J. M. Paredes$^{1}$, B. Marcote$^{3,1}$, K. Iwasawa$^{4}$, J. Mold\'on$^{5,1,6}$, J. Casares$^{7,8,9}$, S. Migliari$^{10}$, X. Paredes-Fortuny$^{1}$}

\affil{$^{1}$Departament de F\'{\i}sica Qu\`antica i Astrof\'{\i}sica, Institut de Ci\`encies del Cosmos (ICCUB), Universitat de Barcelona, IEEC-UB, Mart\'{\i} i Franqu\`es 1, E08028 Barcelona, Spain}
\affil{$^{2}$INAF/IAPS-Roma, I-00133 Roma, Italy}
\affil{$^{3}$ Joint Institute for VLBI ERIC (JIVE), Postbus 2, 7990 AA Dwingeloo, The Netherlands}
\affil{$^{4}$ICREA, Institut de Ci\`encies del Cosmos (ICCUB), Universitat de Barcelona, IEEC-UB, Mart\'i i Franqu\`es 1, E-08028 Barcelona, Spain}
\affil{$^{5}$Jodrell Bank Center for Astrophysics, School of Physics and Astronomy, The University of Manchester, Manchester M13 9PL, UK}
\affil{$^{6}$ASTRON, The Netherlands Institute for Radio Astronomy, Postbus 2, 7990 AA, Dwingeloo, The Netherlands}
\affil{$^{7}$Instituto de Astrof\'isica de Canarias, E-38200 La Laguna, Tenerife, Spain}
\affil{$^{8}$Departamento de Astrof\'isica, Universidad de La Laguna, Avda. Astrof\'isico Francisco S\'anchez s/n, E-38271 La Laguna, Tenerife, Spain}
\affil{$^{9}$Department of Physics, Astrophysics, University of Oxford, Denys Wilkinson Building, Keble Road, Oxford OX1 3RH, UK}
\affil{$^{10}$European Space Astronomy Centre, Apartado/P.O. Box 78, Villanueva
de la Canada, E-28691 Madrid, Spain}

\begin{abstract}
\object{MWC~656} is the first known Be/black hole (BH) binary system. Be/BH binaries are important in the context of binary system evolution and sources of detectable gravitational waves because they are possible precursors of coalescing neutron star/BH binaries. X-ray observations conducted in 2013 revealed that MWC~656 is a quiescent high-mass X-ray binary (HMXB), opening the possibility to explore X-ray/radio correlations and the accretion/ejection coupling down to low luminosities for BH HMXBs. Here we report on a deep joint {\it Chandra}/VLA observation of MWC~656 (and contemporaneous optical data) conducted in 2015 July that has allowed us to unambiguously identify the X-ray counterpart of the source. The X-ray spectrum can be fitted with a power law with $\Gamma\sim2$, providing a flux of $\simeq4\times10^{-15}$~erg~cm$^{-2}$~s$^{-1}$ in the 0.5--8~keV energy range and a luminosity of $L_{\rm X}\simeq3\times10^{30}$~erg~s$^{-1}$ at a 2.6~kpc distance. For a 5~M$_\odot$ BH this translates into $\simeq5\times10^{-9}$~$L_{\rm Edd}$. These results imply that MWC~656 is about 7 times fainter in X-rays than it was two years before and reaches the faintest X-ray luminosities ever detected in stellar-mass BHs. The radio data provide a detection with a peak flux density of $3.5\pm1.1$~$\mu$Jy~beam$^{-1}$. The obtained X-ray/radio luminosities for this quiescent BH HMXB are fully compatible with those of the X-ray/radio correlations derived from quiescent BH low-mass X-ray binaries. These results show that the accretion/ejection coupling in stellar-mass BHs is independent of the nature of the donor star.
\end{abstract}


\section{Introduction} \label{intro}

X-ray observations of stellar-mass black holes (BHs) in binary systems have allowed for the study of the properties of the accretion disks that surround them, as well as their changes over time (see \citealt{remillard06} for a review). Radio observations of these systems have allowed us to gain knowledge of the ejection processes in the form of relativistic jets \citep{mirabel99,fender01,fender16}. Simultaneous observations have revealed the existence of non-linear correlations between the X-ray and radio luminosities during the so-called low/hard and quiescent states that are consistent with scale-invariant jet models \citep{corbel03,gallo03,markoff03,fender04,fender-belloni04,corbel13,gallo14}. However, these studies have mainly been conducted for BHs in low-mass X-ray binaries (LMXBs) because the only confirmed BH in a high-mass X-ray binary (HMXB) in the Galaxy before the discovery of MWC~656, namely \object{Cygnus~X-1}, always displays a high luminosity and does not allow us to trace the correlation (see, e.g., \citealt{zdziarski12}).

\object{MWC~656} is the first binary system containing a BH in orbit around a Be star \citep{casares14}. Its discovery was triggered by the detection, with {\it AGILE}, of the transient gamma-ray source \object{AGL~J2241+4454} \citep{lucarelli10}, undetected with {\it Fermi}/LAT \citep{alexander15} but recently reported to show recurrent activity with hints of long-term variability with {\it AGILE} \citep{munar-adrover16}. The binary nature of MWC~656 (a bright Be star also named \object{HD~215227}) was suggested by a 60.37~d periodicity in optical photometry \citep{williams10,paredes-fortuny12}, and later confirmed through radial velocity studies \citep{casares12}. A detailed spectroscopic analysis revealed the presence of a \ion{He}{2} emission line at $\lambda$4686~\AA\ that could only be produced in an accretion disk around the invisible companion to the Be star. A re-analysis of radial velocity data using an \ion{Fe}{2} emission line from the Be circumstellar disk and the \ion{He}{2} emission line from the accretion disk, together with an improved spectral type classification of B1.5-B2\,III for the Be star, provided a mass for the compact object in the range of 3.8--6.9~M$_\odot$, thus confirming its BH nature \citep{casares14}.

The discovery of the first Be/BH system is relevant in the context of binary system evolution. Although it partially solves the problem of the absence of Be/BH systems \citep{belczynski09}, the discovery of MWC~656 is observationally biased by the lack of a bright X-ray counterpart, suggesting that there might be a large number of hidden Be/BH systems. In addition, simulations have shown that it is very difficult to form Be/BH binaries with properties similar to those of MWC~656, although they can evolve into close NS/BH systems that will merge on timescales of a few Gyr and produce gravitational waves detectable with advanced LIGO/Virgo in nearby galaxies \citep{grudzinska15}.

\cite{munar-adrover14} conducted a 14-ks {\it XMM-Newton} observation of MWC~656 on 2013 June 4 that provided the detection between 0.3 and 5.5~keV of a faint source coincident at the 2.4$\sigma$ level with MWC~656. This revealed that MWC~656 was a new HMXB. The spectrum analysis required a model fit with two components, a blackbody plus a power law, the latter dominating above $\simeq$0.8~keV. The thermal emission was proposed to arise from the wind of the Be star, while the non-thermal emission arose from the vicinity of the BH. The observed non-thermal luminosity was $L_{\rm X}=(1.6^{+1.0}_{-0.9})\times10^{31}$~erg~s$^{-1}$, or $(3.1\pm2.3)\times10^{-8} L_{\rm Edd}$ for a source distance of $2.6\pm0.6$~kpc, indicating a BH in deep quiescence (see \citealt{plotkin13} for reference).

Radio observations with different interferometers, frequencies, and epochs have provided 3$\sigma$ upper limits to the flux density as low as 30~$\mu$Jy \citep{moldon12,marcote15}. Recently, \cite{dzib15} have reported a detection with VLA at a peak flux density of $10\pm3$~$\mu$Jy~beam$^{-1}$. Observations at TeV energies with the MAGIC telescopes have only led to upper limits \citep{aleksic15}. For a detailed review on MWC~656 see \cite{ribo15}.

In this Letter, we report on a deep joint {\it Chandra}/VLA observation (and contemporaneous optical data) of the first Be/BH binary MWC~656 that has allowed us to: 1)~resolve the {\it XMM-Newton} detection into two sources, one of them unambiguously associated with MWC~656; 2)~find that it is significantly fainter than two years before and reaches the faintest quiescent luminosities detected in stellar-mass BHs; and 3)~detect a faint radio counterpart and find that the obtained X-ray/radio luminosities for this quiescent BH HMXB are fully compatible with those of the X-ray/radio correlations derived from quiescent BH LMXBs at the low-luminosity end.

\section{Observations, Data Analysis, and Results} \label{obs}

\subsection{X-rays} \label{x-rays}

MWC~656 was observed with the {\it Chandra X-ray Observatory} ACIS-S camera using the {\tt VFAINT} data mode during $\sim$60~ks, from 2015 July 24 at 21:03~UT to July 25 at 14:29~UT (PI: M. Rib\'o; \dataset [ADS/Sa.CXO#obs/16753] {Chandra ObsId 16753}). The observation covered the orbital phase range 0.01--0.02 (using JD$_0=2453243.3$ from \citealt{williams10}). The data were analyzed using the Chandra Interactive Analysis of Observations (CIAO) software, v4.7\footnote{\tt http://cxc.harvard.edu/ciao/index.html} \citep{fruscione06}. The event files were reprocessed by means of the {\tt chandra\_repro} script. No flares were detected in the background light curve, yielding an effective exposure time of 59.1~ks. The data analysis was performed in the 0.5--8 keV energy band at which {\it Chandra} presents higher sensitivity. 

\begin{figure}[t!]
\resizebox{\hsize}{!}{\includegraphics{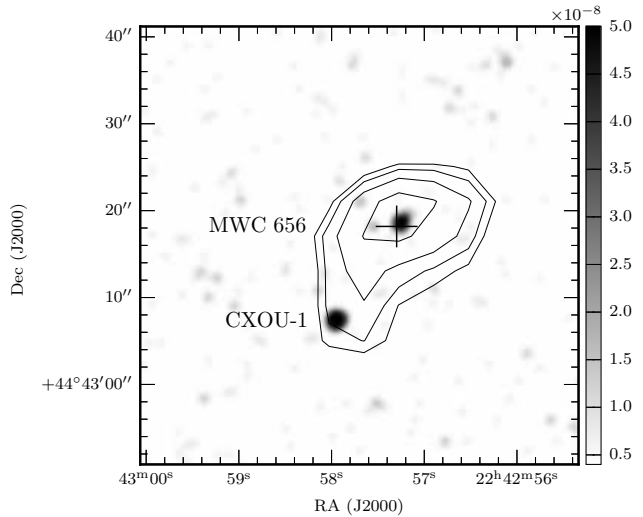}}  
\caption{{\it Chandra} ACIS-S image of $50\arcsec$$\times50\arcsec$ at the position of MWC~656 (indicated by the cross) in the 0.5--8.0 keV energy band smoothed using a Gaussian interpolation with a $3\arcsec$ kernel. The flux scale appears on the right of the image. Contours correspond to the previous {\it XMM-Newton} EPIC-pn image from \cite{munar-adrover14}.
\label{fig:chandra_image}
}
\end{figure}

We show the {\it Chandra} X-ray image of the field around MWC~656 in Figure~\ref{fig:chandra_image}, together with the contours of the previous {\it XMM-Newton} image reported in \cite{munar-adrover14}. Inspection of the X-ray image superimposing optical positions from the USNO-B1.0 catalog \citep{monet03} reveals a systematic offset. Using three common sources we have applied a correction of RA~$=+0{\rlap.}^{\prime\prime}5$ and Dec~$=+0{\rlap.}^{\prime\prime}1$ to the {\it Chandra} data (and of RA~$=+2{\rlap.}^{\prime\prime}4$ and Dec~$=+4{\rlap.}^{\prime\prime}9$ to the {\it XMM-Newton} data using the only common source). The better angular resolution of {\it Chandra} has allowed us to resolve the {\it XMM-Newton} source into two sources. We have used the {\tt wavdetect} algorithm to obtain the significance and positions of both sources. MWC~656 is detected at the 8.5$\sigma$ confidence level (c.l.), with equatorial coordinates (after correction) RA~$=22^{\rm h}42^{\rm m}57{\rlap.}^{\rm s}27 \pm 0{\rlap.}^{\rm s}01$, Dec~$=44^{\circ}43^{\prime}18{\rlap.}^{\prime\prime}5 \pm 0{\rlap.}^{\prime\prime}1$ (uncertainties are statistical only and at 1$\sigma$ c.l.). The position of MWC~656 from {\it Gaia} data at the epoch of the {\it Chandra} observation (epoch=2015.56) following \cite{gaia16} and \cite{gaiacat16} is: RA~$=22^{\rm h}42^{\rm m}57{\rlap.}^{\rm s}29718 \pm 0{\rlap.}^{\rm s}00002$, Dec~$=44^{\circ}43^{\prime}18{\rlap.}^{\prime\prime}2099 \pm 0{\rlap.}^{\prime\prime}0002$. The offset between both positions is $0{\rlap.}^{\prime\prime}4~\pm~0{\rlap.}^{\prime\prime}1_{\rm stat}~\pm~0{\rlap.}^{\prime\prime}1_{\rm syst}$, and the {\it Chandra} source is thus the counterpart of MWC~656.

The new source, located $\sim$13${\rlap.}^{\prime\prime}5$ to the southeast of MWC~656, is detected at a 15.9$\sigma$ c.l. and has equatorial coordinates (after correction) RA~$=22^{\rm h}42^{\rm m}57{\rlap.}^{\rm s}969 \pm 0{\rlap.}^{\rm s}005$, Dec~$=44^{\circ}43^{\prime}07{\rlap.}^{\prime\prime}34 \pm 0{\rlap.}^{\prime\prime}05$, and thus is named \object{CXOU~J224257.9+444307} (CXOU-1 from now on). The {\it Chandra} image reveals that MWC~656 is fainter than CXOU-1 (see below), while the shape of the {\it XMM-Newton} contours reveals that in 2013 June MWC~656 was brighter than CXOU-1.

\begin{figure*}[]
\plottwo{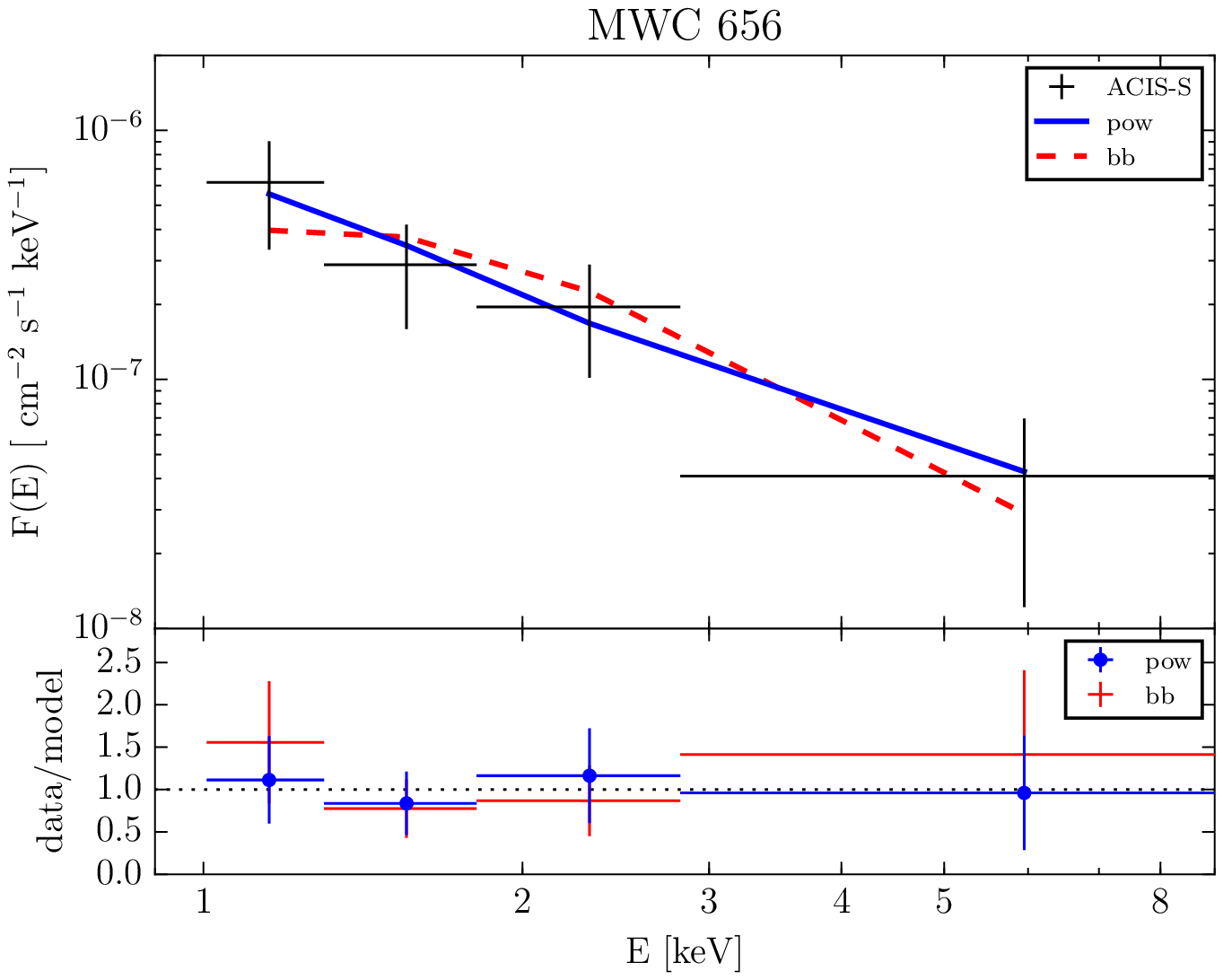}{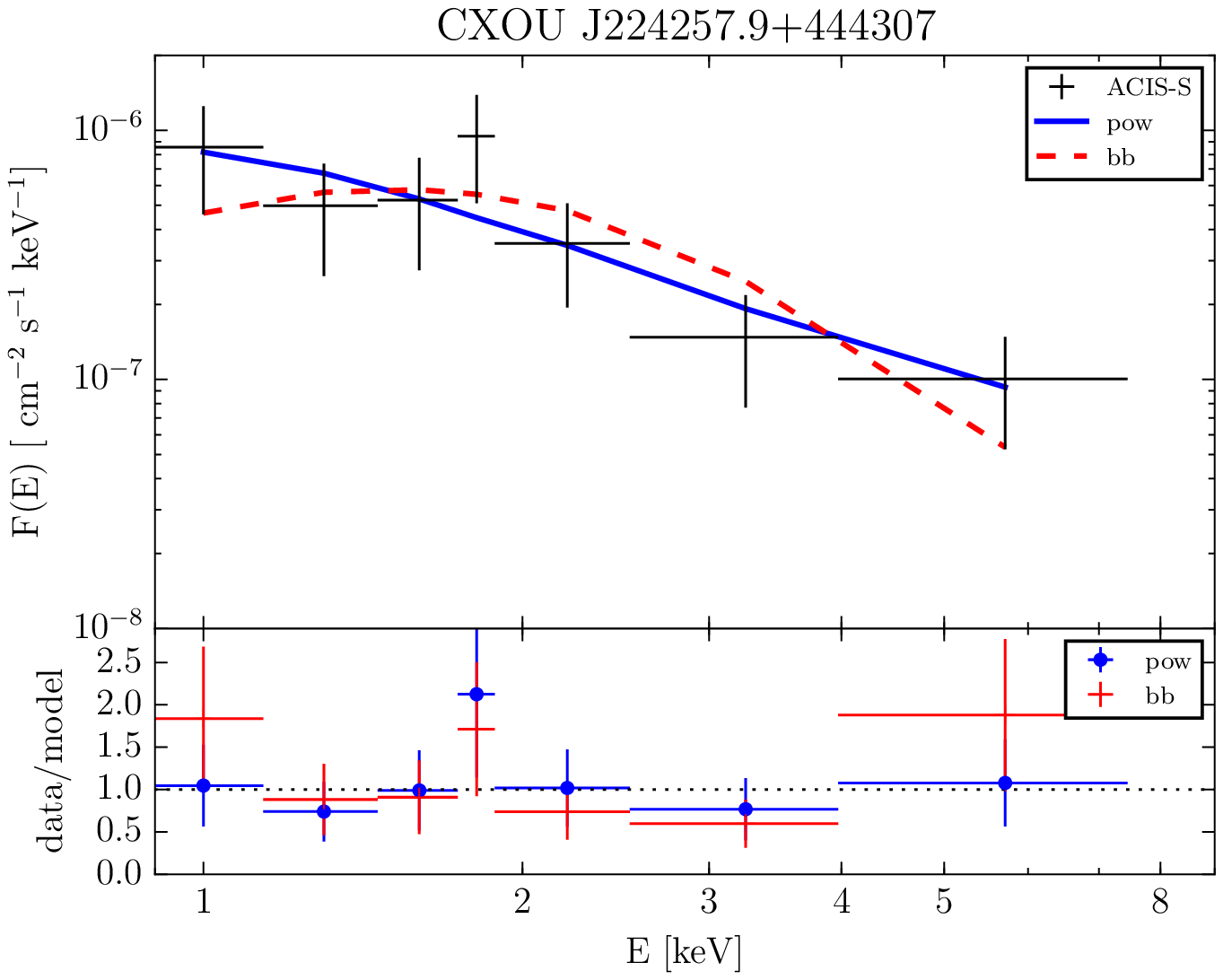}
\caption{Left: MWC~656 {\it Chandra} ACIS-S spectrum in the 0.5--8.0 keV energy range (crosses) overplotted with the fitted absorbed power-law model (blue solid line) and blackbody model (red dashed line). The lower panel illustrates the ratio between the observational data and the corresponding model (blue crosses with circles for the power-law model and red crosses for the blackbody model).
Right: the same figure for the new source CXOU~J224257.9+444307.
\label{hd_spec}
}
\label{fig:spectra}
\end{figure*}

\begin{table*}[]
\begin{center}
\caption{Summary of spectral fit results \label{table:spectra}}
\begin{tabular}{lccccccc}
\hline
\hline
Source & \multicolumn{3}{c}{Power law} & & \multicolumn{3}{c}{Blackbody} \\
\cline{2-4}
\cline{6-8}
  & $\Gamma$ & $F(0.5$--$8~{\rm keV})$             & $C$-statistic & & $k_{\rm B}T$ & $F(0.5$--$8~{\rm keV})$             & $C$-statistic \\
  &          & $10^{-15}$ [erg cm$^{-2}$ s$^{-1}$] &               & & [keV]        & $10^{-15}$ [erg cm$^{-2}$ s$^{-1}$] & \\  
\hline
MWC~656               & $2.2^{+1.3}_{-0.9}$ & $4.1^{+2.3}_{-1.5}$ & 0.3 & & $0.6\pm0.3$ & $2.7^{+1.5}_{-1.1}$ & 3.0  \\
CXOU~J224257.9+444307 & $1.8\pm0.6$         & $8.0^{+2.8}_{-2.1}$ & 4.4 & & $0.7\pm0.2$ & $6.0^{+2.3}_{-1.7}$ & 12.5 \\
\hline
\end{tabular}
\tablecomments{Uncertainties are at 68\% c.l. Flux values are unabsorbed.}
\end{center}
\end{table*}

With the use of the {\tt srcflux} script we extracted net counts for MWC~656 ($22.1\pm4.9$ at 90\% c.l.) and CXOU-1 ($38.7\pm6.3$), taking into account the background and the point-spread function of the instrument. These measurements were obtained from a region centered at the peak of the X-ray emission (found by {\tt wavdetect}) of each source with a radius of 3\arcsec, while background counts were obtained from a nearby circular region with a 10\arcsec\ radius, avoiding contamination from nearby sources, always within the 0.5--8.0~keV energy band. 

We used XSPEC version 12.8.2 \citep{arnaud96} to perform a spectral analysis of both sources. The same regions for source and background used to extract net counts were considered. We created response files in order to take into account the energy-dependent behavior of the instrument. We fitted the spectra with two distinct models, an absorbed power law and an absorbed blackbody, always using a fixed column density of $N_{\rm H}=1.8\times10^{21}$~cm$^{-2}$ (see \citealt{munar-adrover14}). The low number of counts required the use of the {\tt cstat} statistic within XSPEC to estimate the best-fit parameters and their associated uncertainties \citep{cash79}.

We show the spectra of MWC~656 and CXOU-1, together with the fitted models, in Figure~\ref{fig:spectra}. The results of the fits are listed in Table~\ref{table:spectra}. Although both sources can be fitted with either model, the obtained $C$-statistics reveal that the power-law model is preferred since blackbody models result in an excess at both low and high energies. The obtained fluxes for the power-law models in the 0.5--8~keV range are $4.1^{+2.3}_{-1.5}\times10^{-15}$~erg~cm$^{-2}$~s$^{-1}$ for MWC~656 and $8.0^{+2.8}_{-2.1}\times10^{-15}$~erg~cm$^{-2}$~s$^{-1}$ for CXOU-1. Although CXOU-1 appears brighter in Figure~\ref{fig:chandra_image}, the fluxes obtained by the spectral analysis are compatible at 1$\sigma$ c.l. The power-law fit for MWC~656 provides a photon index of $2.2^{+1.3}_{-0.9}$, while for CXOU-1 we obtain $1.8\pm0.6$, thus fully compatible at 1$\sigma$ c.l. The conclusion from the analysis of the {\it Chandra} data is that both sources have a similar spectrum, while the X-ray flux of MWC~656 is approximately half of the one measured for CXOU-1. There is no reported evidence for this source in the literature and the low number of photons prevents us from clearly establishing its nature.

We have tested if the {\it Chandra} data for the two sources produces a spectrum that needs a two-component model. We have combined the spectra of the two sources and find that the data can be reproduced with a single power law with a photon index of $2.1\pm0.5$ ($C$-statistic 3.7). Therefore, the {\it Chandra} data do not support the two-component model reported in \cite{munar-adrover14}.

\subsection{Radio} \label{radio}

MWC~656 was observed with the Karl G. Jansky Very Large Array (VLA) of the National Radio Astronomy Observatory (NRAO) on 2015 July 25 from 05:00 to 11:00~UTC (during the {\it Chandra} observation) in the A configuration. The observation was conducted at 10~GHz, with full circular polarization, using 32 spectral windows with 64 channels of 2~MHz bandwidth each, providing a total bandwidth of 4~GHz. The amplitude calibrator was 3C~48. The phase calibrator J2255+4202 was observed in 1:20-min runs interleaved with 6-min runs on the target source.

The calibration was conducted using version 4.5.0 of the CASA package\footnote{\url{http://casa.nrao.edu}} of NRAO \citep{mcmullin07}. The data have been reduced using standard amplitude and phase calibration steps. We imaged the data using the {\tt clean} procedure with a natural weighting, obtaining a synthesized beam of $0.35 \times 0.19~\mathrm{arcsec^2}$ in a Position Angle (P.A.) of 47$^{\circ}$ (north to east). Due to the proximity of a bright quasar affecting the field of MWC~656 (see \citealt{marcote15}), we imaged a region of a few arcmin including this quasar. We conducted a multi-scale clean, considering the usual point-like components and extended components with sizes one to three times the synthesized beam (see \citealt{cornwell08} and \citealt{rich08} for details). To check the reliability of the obtained results we also analyzed the data using standard clean, different weighting schemes, and different time intervals, which always led to compatible results within uncertainties.

\begin{figure}[t!]
\resizebox{\hsize}{!}{\includegraphics{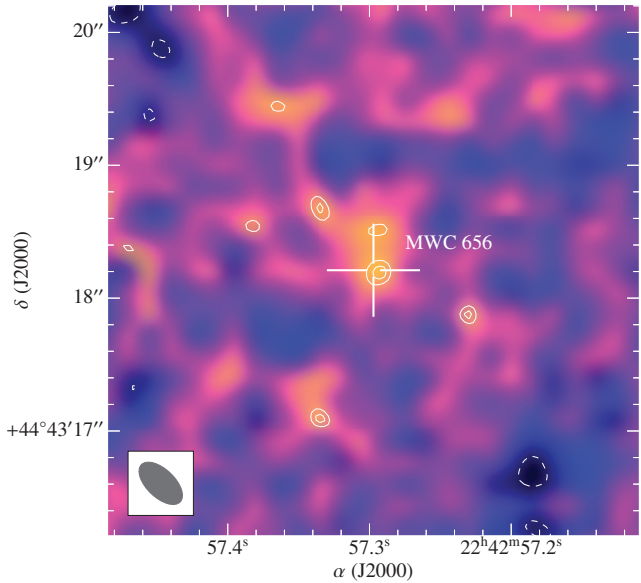}}
\caption{VLA image of $4\arcsec$$\times4\arcsec$ centered at the optical position of MWC~656 (indicated by the open cross) obtained at 10~GHz with the VLA in its A configuration. The synthesized beam size is of $0.35 \times 0.19~\mathrm{arcsec^2}$ in P.A. of 47$^{\circ}$. Contours represent $-$2.5, 2.5, and 3 times the rms noise of 1.1~$\mu$Jy~beam$^{-1}$.
\label{fig:vla_image}
}
\end{figure}

The obtained image is shown in Figure~\ref{fig:vla_image}. A faint radio source with a peak flux density of $3.5\pm1.1$~$\mu$Jy~beam$^{-1}$ is detected at the optical position of MWC~656 within uncertainties. Although this is a marginal detection, there are several facts that support the reality of the radio source as the counterpart of MWC~656: 1) it is the brightest radio source in the image, 2) it is the only one located within the {\it Chandra} X-ray source position uncertainty, 3) it is fully compatible with the optical position, and 4) there is a former radio detection of MWC~656 \citep{dzib15}.

Inspection at the position of CXOU-1 reveals no radio counterpart, with a 3$\sigma$ flux density upper limit of 3.2~$\mu$Jy for a point-like source.

\subsection{Optical} \label{optical}

MWC~656 was observed with different optical facilities to obtain contemporaneous photometric and spectroscopic data. Photometric observations were conducted with a passband optical filter at the 0.5~m robotic Telescope Fabra Roa Montsec (TFRM, \citealt{fors13}) on the nights from 2015 July 22 to 27 (except the 24), and with $BVRI_{\rm c}$ filters at the 0.8~m robotic Joan Or\'o Telescope (TJO, \citealt{colome10}) from 2015 July 28 onwards. Both telescopes are located at Observatori Astron\`omic del Montsec (OAdM, Sant Esteve de la Sarga, Catalonia). While detailed results will be published elsewhere, during the {\it Chandra}/VLA observation MWC~656 displayed photometry compatible with its already known behavior \citep{paredes-fortuny12,paredes-fortuny16}. Inspection of focused images taken on 2015 November 12 reveals no optical counterpart for CXOU-1, implying $I_{\rm c}>18.5$~mag.

MWC~656 was also observed with the fiber-fed STELLA  Echelle Spectrograph (SES) of the 1.2~m robotic STELLA-I (ST) optical telescope \citep{strassmeier04} at the Observatorio del Teide (OT, Tenerife, Spain) on the nights of 21--25 July 2015 (except the 22). The setup and the data reduction were the same as in \cite{aleksic15}. The spectra show the presence of the double peaked \ion{He}{2} $\lambda$4686 emission line with an equivalent width comparable to that reported in \cite{casares14}. We also detect other emission lines, mainly H$\alpha$, H$\beta$ and weak \ion{Fe}{2} lines with comparable strengths to those measured by \cite{casares12}. Therefore, MWC~656 was in a similar optical state as in past observations.

\section{Discussion and conclusions} \label{discussion}

The {\it Chandra} observation of MWC~656 reported here has revealed that the {\it XMM-Newton} source presented in \cite{munar-adrover14} was in fact the superposition of two sources: MWC~656 and the new source CXOU~J224257.9+444307 (CXOU-1). Here MWC~656 is significantly fainter than it was in the {\it XMM-Newton} observation. The VLA observation of MWC~656 has led to the detection of a faint source. This is the first simultaneous X-ray/radio detection of the first Be/BH binary system. Here we discuss the implications of these detections.

First, it is worth noting that during the {\it Chandra}/VLA observation MWC~656 was showing optical photometric and spectroscopic properties compatible with those reported in previous works, indicating that there was no significant change in either the Be star and its circumstellar disk or the outer part of the BH accretion disk (\ion{He}{2} emission line). In addition, inspection of the {\it MAXI} data \citep{matsuoka09} in the direction of AGL~J2241+4454 reveals no X-ray emission up to now (2009 August to 2016 October). Therefore, MWC~656 appears to have been in a long quiescent X-ray state, at least during the last 7 years.

\begin{figure*}[t!]
\begin{center}
\resizebox{0.8\hsize}{!}{\includegraphics{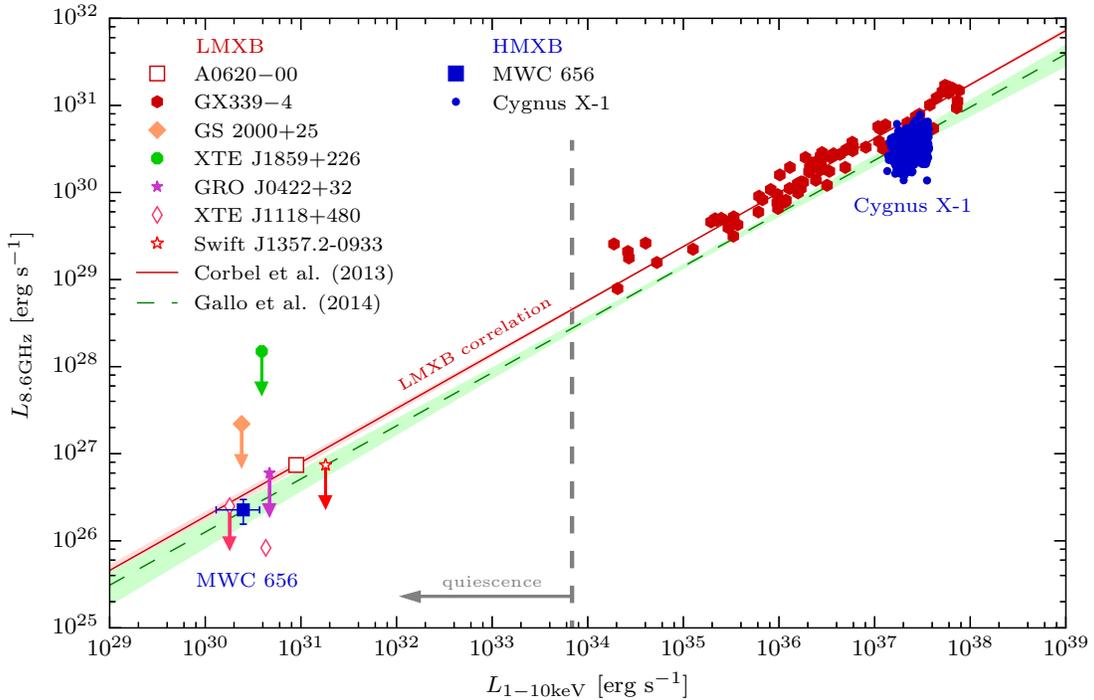}}  
\caption{Radio vs. X-ray luminosity diagram including the position of 
MWC~656 (with 1$\sigma$ uncertainties) obtained with the {\it Chandra}/VLA observation reported in this work and the positions of a few additional significant sources from simultaneous observations. We plot the X-ray/radio correlations for BH LMXBs from \citeauthor{corbel13}\ (\citeyear{corbel13}; solid red line with a light red shadow) and from \citeauthor{gallo14}\ (\citeyear{gallo14}; dashed green line with a light green shadow). To display the luminosity range of LMXBs that follow the correlation we show data on the BH LMXBs GX~339$-$4 (red hexagons), as well as the two faintest sources ever detected in radio and X-rays simultaneously: A0620$-$00 (open red square; \citealt{gallo06}) and XTE~J1118+480 (light red diamond; \citealt{gallo14}). The faintest X-ray detections of LMXBs without radio counterparts have also been plotted for reference. The small blue dots indicate the region of the parameter space where Cygnus~X-1 has been detected in the low/hard state \citep{gallo12}. The hard-state radio-quiet outliers from the correlation have not been plotted. The gray dashed line separates the quiescent state region (left) and the other states (right) according to \cite{plotkin13}. 
\label{fig:correlation}
}
\end{center}
\end{figure*}

The X-ray flux of MWC~656 provided by {\it Chandra} in 2015 July is $4.1^{+2.3}_{-1.5}\times10^{-15}$~erg~cm$^{-2}$~s$^{-1}$ in the 0.5--8~keV energy range. For comparison, the previous X-ray observation conducted with {\it XMM-Newton} in 2013 June provided a total flux of $4.6^{+1.3}_{-1.1}\times10^{-14}$~erg~cm$^{-2}$~s$^{-1}$ in the 0.3--5.5~keV energy range, which translates into $3.5\pm0.9\times10^{-14}$~erg~cm$^{-2}$~s$^{-1}$ in the 0.5--8~keV energy range. This flux is approximately one order of magnitude larger than the one measured for MWC~656 with {\it Chandra}. However, the {\it XMM-Newton} source was the superposition of two sources. Assuming that CXOU-1 had a constant flux of $\sim8.0\times10^{-15}$~erg~cm$^{-2}$~s$^{-1}$ in the 0.5--8~keV energy range (as determined from the {\it Chandra} data), the corresponding flux of MWC~656 at the time of the {\it XMM-Newton} observation was $\sim2.7\times10^{-14}$~erg~cm$^{-2}$~s$^{-1}$ in the same energy range. This would indicate a decrease of a factor of $\sim$7 in the X-ray flux of MWC~656 between 2013 June and 2015 July.

Considering a distance of $2.6\pm0.6$~kpc to MWC~656 \citep{casares14}, the {\it Chandra} X-ray luminosity of the source in the 0.5--8~keV energy range is $L_{\rm X}=(3.3^{+2.4}_{-2.0})\times10^{30}$~erg~s$^{-1}$ (including the flux and distance uncertainties). Considering a BH mass in the range of 3.8--6.9~M$_\odot$ this translates into $(4.9^{+3.9}_{-3.2})\times10^{-9}$~$L_{\rm Edd}$. In the 1--10~keV energy range these values are: $L_{\rm X}=(2.5^{+2.6}_{-1.7})\times10^{30}$~erg~s$^{-1}$ = $(3.7^{+4.0}_{-2.7})\times10^{-9}$~$L_{\rm Edd}$. These luminosities and the obtained photon index are fully compatible with a BH in deep quiescence \citep{plotkin13}.

The simultaneous VLA observation of MWC~656 has yielded a detection with a peak flux density of $3.5\pm1.1$~$\mu$Jy~beam$^{-1}$. The accreting BH in MWC~656 could easily produce it through synchrotron emitting electrons in a jet, as seen in many X-ray binaries (e.g., \citealt{fender16} and references therein). In contrast, the production of such flux density by gyro-synchrotron radiation in the magnetic field of the Be star would require too high magnetic fields\footnote{We note the following: 1) the rapid rotation of Be stars prevents the existence of high magnetic fields in these objects; and 2) magnetism is less present in massive binaries than in isolated massive stars \citep{schoeller14,neiner15}.} combined with relatively high electron densities above the energy threshold of 10~keV \citep{dulk85,guedel02}. Therefore, in what follows we consider that the detected radio emission in MWC~656 has a synchrotron origin in a jet.

We show in Figure~\ref{fig:correlation} the radio versus X-ray luminosity diagram for BH X-ray binaries. The simultaneous {\it Chandra}/VLA observation of MWC~656 allows us to place the source within the diagram in a reliable way\footnote{The position of MWC~656 in Figure~3 of \cite{dzib15} is based on their VLA data obtained in 2015 Feb-Apr and the non-simultaneous {\it XMM-Newton} data obtained in 2013 June. In addition, our {\it Chandra} observation shows that the {\it XMM-Newton} flux corresponds to two sources and that the X-ray luminosity of MWC~656 is significantly variable.}. MWC~656 is one of the faintest stellar-mass BHs ever detected in X-rays, together with \object{XTE~J1118+480}, \object{GS~2000+25}, \object{XTE~J1859+226}, \object{GRO~J0422+32}, \object{A0620$-$00}, and \object{Swift~J1357.2$-$0933}, \citep{gallo14,miller-jones11,gallo03,gallo06,plotkin16}, and the faintest one in X-rays also detected in radio. The obtained luminosities for this quiescent BH HMXB are fully compatible with those of the X-ray/radio correlations derived from quiescent BH LMXBs at the low-luminosity end. Together with Cygnus~X-1 at the high-luminosity end, it is now clear, for the first time, that the accretion/ejection coupling in stellar-mass BHs is independent of the nature of the donor star.

Finally, given the X-ray variability found in MWC~656, future simultaneous X-ray/radio observations should allow us to trace the motion of the source in the radio versus X-ray luminosity diagram, and directly check the slope of the correlation at the low-luminosity end for the first time in HMXBs.

\acknowledgments

We thank the anonymous reviewer for providing suggestions that helped to improve the original version of the manuscript.
We thank S.\ Corbel for providing the data for LMXBs and E.\ Gotthelf for useful discussions on the {\it Chandra} astrometry.
The scientific results reported in this article are based to a significant degree on observations made by the {\it Chandra X-ray Observatory}.
This research has made use of software provided by the {\it Chandra} X-ray Center (CXC) in the application package CIAO.
This research has made use of the XSPEC software.
The National Radio Astronomy Observatory is a facility of the National Science Foundation operated under cooperative agreement by Associated Universities, Inc.
The Common Astronomy Software Applications, CASA, is a software produced and maintained by the NRAO.
The authors acknowledge support of the TFRM team for preparing and carrying out the optical photometric observations.
The Joan Or\'o Telescope (TJO) of the Montsec Astronomical Observatory (OAdM) is owned by the Catalan Government and operated by the Institute for Space Studies of Catalonia (IEEC).
This work has made use of data from the European Space Agency (ESA) mission {\it Gaia} (\url{http://www.cosmos.esa.int/gaia}), processed by the {\it Gaia} Data Processing and Analysis Consortium (DPAC, \url{http://www.cosmos.esa.int/web/gaia/dpac/consortium}). Funding for the DPAC has been provided by national institutions, in particular the institutions participating in the {\it Gaia} Multilateral Agreement.
This research has made use of the {\it MAXI} data provided by RIKEN, JAXA, and the {\it MAXI} team.
This research has made use of NASA's Astrophysics Data System; the SIMBAD database and the VizieR catalogue access tool \citep{ochsenbein00}, operated at CDS, Strasbourg, France; and Astropy, a community-developed core Python package for Astronomy \citep{astropy13}.
We acknowledge support by the Spanish Ministerio de Econom\'ia y Competitividad (MINECO/FEDER, UE) under grants AYA2013-47447-C3-1-P, AYA2013-47447-C3-2-P, AYA2013-42627, AYA2016-76012-C3-1-P, FPA2015-69210-C6-2-R, MDM-2014-0369 of ICCUB (Unidad de Excelencia `Mar\'ia de Maeztu'), SEV-2015-0548 of IAC (Centro de Excelencia `Severo Ochoa'), and the Catalan DEC grant 2014 SGR 86.
P.M.A. acknowledges partial support through the ASI grant no.\ I/028/12/0.
J.C. acknowledges support by the Leverhulme Trust through the Visiting Professorship Grant VP2-2015-046.

{\it Facilities:} 
\facility{CXO(ACIS)}, \facility{VLA}.


\end{document}